\documentclass[12pt,a4paper]{article}%
\usepackage{amssymb}
\usepackage{amsmath}%
\usepackage{amsfonts}%
\usepackage{graphicx}
\begin{document}

\title{\textbf{Energy Spectrum of a 2D Dirac Oscillator in the Presence of the
Aharonov-Bohm Effect}}
\author{N. Ferkous \ and \ A. Bounames \thanks{E-mail: a.bounames@gmail.com}\\
\textit{{Laboratory of Theoretical Physics, University of Jijel},}\\\textit{BP 98, Ouled Aissa, Jijel 18000, Algeria}}
\date{}
\maketitle

\begin{abstract}
We determine the energy spectrum and the corresponding eigenfunctions of a 2D
Dirac oscillator in the presence of Aharonov-Bohm (AB) effect . It is shown
that the energy spectrum depends on the spin of particle and the AB magnetic
flux parameter. Finally, when the irregular solution occurs it is shown that
the energy takes particular values. The nonrelativistic limit is also
considered.\newline\ 

PACS \ 03.65, 03.65 Pm, 03.65 Bz

\end{abstract}


\setcounter{MaxMatrixCols}{10}


\newpage

\section{Introduction}

The Dirac oscillator was introduced for the first time by Moshinsky and
Szczepaniak \cite{Moushinsky}. The basic idea is that the momentum
$\mathbf{p}$ is replaced in the free Dirac equation by $\mathbf{p}%
\rightarrow\mathbf{p}-iM\omega\beta\mathbf{r},$ where $\mathbf{r}$ is the
position vector, $M$\ \ is the mass of particle and $\omega$ the frequency of
the oscillator. The Dirac oscillator has attracted much attention and has
various physical applications \cite{MoushinskyB,rozmej,pacheco}, particularly
in semiconductor physics \cite{Vil2}. In the nonrelativistic limit, the Dirac
oscillator becomes an harmonic oscillator with strong spin-orbit coupling term.

Recently, the 2D electron systems has become an active research subject due to
the advances in nanofabrication technology like quantum wells, quantum wires,
quantum dots, quantum Hall effect and high superconductivity \cite{Vil2,Vil3}.
Currently, the 2D parabolic potential $ V=\frac{1}{2}M \omega^2 r^2 $ 
is often used to describe confined 2D systems in nonrelativistic case. For
relativistic case, such systems are described using the 2D Dirac oscillator
\cite{Vil2}.

On the other hand, the Aharonov-Bohm (AB) effect \cite{aharo} leads to a
number of remarkable interference phenomena in mesoscopic systems
\cite{yasha}, and has been used also to study the 2D model in both
superconductivity and particle theory \cite{HAGEN1}. In this article, in order
to study a bound state version of the AB effect, let us consider the 2D Dirac
oscillator of relativistic particle in the presence of the AB effect. The
associate magnetic field is assumed to be confined to a filament of
vanishingly small radius and perpendicular to the plane where the particles
are confined to move. We study the dependence of the bound states energy and
eigenfunctions on the AB magnetic flux parameter and the spin of particle.

This paper is organized as follows. In Section 2, we solve the Dirac
oscillator in polar coordinates with an additional delta-function interaction
due to the AB effect. In Section 3, we analyze and discuss the energy
spectrum and its degeneracy. The nonrelativistic limit is also considered.
Along the paper we adopt the natural units, $\hslash=1,$ $c=1$ .

\bigskip

\section{ Solution of the Dirac equation}

In the beginning, one writes the stationary Dirac equation for a particle of mass
$M$ in terms of two-component spinors $\psi$%
\begin{equation}
E\psi=\left(  \beta\mathbf{\gamma\pi}+\beta M\right)  \psi,\label{1}%
\end{equation}
where \ $\mathbf{\pi}=\mathbf{p}-e\mathbf{A}$ is the minimally coupling and
$\mathbf{A}$ is the vector potential.

The 2D Dirac oscillator is defined by changing the momentum $\mathbf{p}$ in
the Dirac equation \cite{Vil2, Vil1} as
\begin{equation}
\mathbf{p}\rightarrow\mathbf{p}-iM\omega\beta\mathbf{r}.\label{2}%
\end{equation}
The magnetic field $H$, associated to the AB effect, is assumed to be
perpendicular to the plane and confined to a filament of vanishingly small
radius so that the flux \cite{HAGEN1,HAGEN2}

\begin{equation}
\alpha=-e\int_{0}^{\infty}H(r)rdr,\label{3}%
\end{equation}
is finite and nonzero. Since, however, one would ultimately like to pass to
the limit in which the field is confined, $H$ \ must be taken to be of the
form \cite{HAGEN1,HAGEN2}%
\begin{equation}
eH=-\frac{\alpha}{R}\delta(r-R),\label{4}%
\end{equation}
the corresponding form of the\ potential $\mathbf{A}$\ in the Coulomb gauge is
\cite{HAGEN1,HAGEN2}
\begin{equation}
e\mathbf{A}=\left\{
\begin{array}
[c]{c}%
-\frac{\alpha}{r}\mathbf{u}_{\mathbf{\theta}},\text{ \ \ \ \ }r>R,\\
0,\text{ \ \ \ \ \ \ }r<R.
\end{array}
\right.
\end{equation}
Then, the stationary Dirac equation of a 2D Dirac oscillator in the presence
of the AB effect is%
\begin{equation}
E\psi=\left[  \beta\mathbf{\gamma}\left(  \mathbf{\pi}-iM\omega\beta
\mathbf{r}\right)  +\beta M\right]  \psi.\label{6}%
\end{equation}
Since we are using only two component spinors, the matrices $\beta$ and
$\beta\gamma_{i}$\ are\ conveniently defined in terms of the Pauli spin
matrices as%
\begin{equation}
\beta\gamma_{1}=\sigma^{1},\text{ \ }\beta\gamma_{2}=s\sigma^{2},\text{
\ }\beta=\sigma^{3},\text{ }s=\pm1.\label{7}%
\end{equation}
Eq. (\ref{6}) can be rewritten as%
\begin{equation}
\beta\left[  \mathbf{\gamma}\left(  \mathbf{\pi}-iM\omega\beta\mathbf{r}%
\right)  +M-\beta E\right]  \psi=0.\label{8}%
\end{equation}
By applying the following matrix operator
\[
\left[  \mathbf{\gamma}\left(  \mathbf{\pi}-iM\omega\beta\mathbf{r}\right)
-M-\beta E\right]  \beta
\]
to Eq. (\ref{8}), we obtain
\begin{equation}
\left[  \mathbf{\gamma}\left(  \mathbf{\pi}-iM\omega\beta\mathbf{r}\right)
\mathbf{\gamma}\left(  \mathbf{\pi}-iM\omega\beta\mathbf{r}\right)  -M^{2}+E%
{{}^2}%
\right]  \psi=0.\label{8a}%
\end{equation}
The first term in the left hand side can be computed as%
\begin{align*}
\mathbf{\gamma}\left(  \mathbf{\pi}-iM\omega\beta\mathbf{r}\right)
\mathbf{\gamma}\left(  \mathbf{\pi}-iM\omega\beta\mathbf{r}\right)   &
=\mathbf{\gamma\pi\gamma\pi}+2M\omega\beta+2M\omega s(\mathbf{\pi}%
\wedge\mathbf{r})_{z}\text{ }\beta-M%
{{}^2}%
\omega%
{{}^2}%
r%
{{}^2}%
\\
& =\mathbf{\gamma\pi\gamma\pi}+2M\omega\beta-2M\omega s\left[  L_{z}%
+\alpha\right]  \beta-M%
{{}^2}%
\omega%
{{}^2}%
r%
{{}^2}%
.
\end{align*}
Thus, Eq. (\ref{8a}) reads%
\begin{equation}
\left[  \mathbf{\gamma\pi\gamma\pi}-M%
{{}^2}%
\omega%
{{}^2}%
r%
{{}^2}%
+2M\omega\left[  1-s\left(  L_{z}+\alpha\right)  \right]  \beta+E%
{{}^2}%
-M^{2}\right]  \psi=0,\label{9}%
\end{equation}
where
\begin{equation*}
\mathbf{\gamma\pi\gamma\pi} =-\mathbf{\pi}^{2}+es\sigma_{3}H,\qquad 
L_{z} =\frac{1}{i}\frac{\partial}{\partial\theta}.
\end{equation*}
Then, in the coordinates representation, Eq. (\ref{9}) reads
\begin{equation}
\Bigg\{\frac{1}{r}\frac{\partial}{\partial r}r\frac{\partial}{\partial r}%
+\frac{1}{r^{2}}(\frac{\partial}{\partial\theta}+i\alpha)^{2}-M%
{{}^2}%
\omega%
{{}^2}%
r%
{{}^2}%
+E%
{{}^2}%
-M^{2}+2M\omega \left[1-s(\frac{1}{i}\frac{\partial}{\partial\theta}%
+\alpha) \right]\sigma^{3}-\frac{\alpha s}{R}\delta(r-R)\sigma^{3}\Bigg\}\psi=0,
\end{equation}
we note that the last equation contains a spin-orbit coupling term.

Let us define the first component as%
\begin{equation}
\psi_{1}=\overset{+\infty}{\underset{m=-\infty}{\sum}}f_{m}(r)e^{im\theta
},\label{12}%
\end{equation}
where the radial part satisfy the equation%
\begin{equation}
\left[  \frac{d^{2}}{dr^{2}}+\frac{1}{r}\frac{d}{dr}-M%
{{}^2}%
\omega%
{{}^2}%
r%
{{}^2}%
-\frac{\left(  m+\alpha\right)  ^{2}}{r^{2}}+E%
{{}^2}%
-M^{2}+2M\omega \left[1-s\left(  m+\alpha\right) \right]-\frac{\alpha
s}{R}\delta(r-R)\right]  f_{m}(r)=0.\text{ }\label{13}%
\end{equation}
Thus, it yields the following equations for the two regions%
\begin{equation}
\left[  \frac{d^{2}}{dr^{2}}+\frac{1}{r}\frac{d}{dr}-\lambda^{2}r%
{{}^2}%
-\frac{m^{2}}{r^{2}}+k_{in}^{2}\right]  f_{m}(r)=0,\ \ \ r<R,\label{14}%
\end{equation}%
\begin{equation}
\left[  \frac{d^{2}}{dr^{2}}+\frac{1}{r}\frac{d}{dr}-\lambda^{2}r%
{{}^2}%
-\frac{(m+\alpha)^{2}}{r^{2}}+k_{out}^{2}\right]  f_{m}(r)=0\text{ },\text{
\ \ }r>R,\label{15}%
\end{equation}
where%
\begin{equation}
k_{out}^{2}=E%
{{}^2}%
-M^{2}+2M\omega\left[  1-s\left(  m+\alpha\right)  \right]  ,\label{16}%
\end{equation}%
\begin{equation}
k_{in}^{2}=E%
{{}^2}%
-M^{2}+2M\omega\left(  1-sm\right)  ,\text{ \ \ }\lambda^{2}=M%
{{}^2}%
\omega%
{{}^2}%
.\label{17}%
\end{equation}
The general solution of the radial equation (\ref{15}) is given in terms of
the confluent hypergeometric function%
\begin{align}
f_{m}(r)& =\exp(-\frac{M\omega}{2}r^{2})\bigg\{A_{m}r^{\left(m+\alpha
\right)}\text{ \ }_{1}F_{1}\left(  \frac{1}{2}\left[m+\alpha
+1-\frac{k_{out}^{2}}{2M\omega}\right], m+\alpha+1, M\omega r^{2}\right)
\text{ }\nonumber\\
& +B_{m}r^{-\left(m+\alpha\right)}\text{ }_{1}F_{1}\left(\frac{1}%
{2}\left[1-\left(  m+\alpha\right)  -\frac{k_{out}^{2}}{2M\omega}\right]
,1-\left(  m+\alpha\right)  ,M\omega r^{2}\right)  \bigg\},\label{18a}
\end{align}
where $A_{m},$ $B_{m}$ and $C_{m}$ are constants.

This equation can be rewritten as%
\begin{align}
f_{m}(r)  & =\exp(-\frac{M\omega}{2}r^{2})\bigg\{A_{m}r^{\left\vert
m+\alpha\right\vert }\text{ }_{1}F_{1}\left(  \frac{1}{2}\left[  \left\vert
m+\alpha\right\vert +1-\frac{k_{out}^{2}}{2M\omega}\right], \left\vert
m+\alpha\right\vert +1, M\omega r^{2}\right) \nonumber\\
& +B_{m}r^{-\left\vert m+\alpha\right\vert }\text{ \ }_{1}F_{1}\left(
\frac{1}{2}\left[ 1-\left\vert m+\alpha\right\vert -\frac{k_{out}^{2}%
}{2M\omega}\right], 1-\left\vert m+\alpha\right\vert, M\omega r^{2}\right)
\bigg\},\label{19}
\end{align}
and consequently for the two regions we have
\begin{equation}
f_{m}(r)_{in}=C_{m}r^{\left\vert m\text{ }\right\vert }\exp(-\frac{M\omega}%
{2}r^{2})\text{ }_{1}F_{1}\left(  \frac{1}{2}\left[  \left\vert m\right\vert
+1-\frac{k_{in}^{2}}{2M\omega}\right]  ,\left\vert m\right\vert +1,M\omega
r^{2}\right)  ,\label{20}%
\end{equation}

$\bigskip$for $r<R,$ and%
\begin{align}
f_{m}(r)_{out}  & =\exp(-\frac{M\omega}{2}r^{2})\bigg\{A_{m}r^{\left\vert
m+\alpha\right\vert }\text{ }_{1}F_{1}\left(  \frac{1}{2}\left[  \left\vert
m+\alpha\right\vert +1-\frac{k_{out}^{2}}{2M\omega}\right]  ,\left\vert
m+\alpha\right\vert +1,M\omega r^{2}\right) \nonumber\\
& \text{\ \ }+B_{m}M\omega r^{-\left\vert m+\alpha\right\vert }\text{ }%
_{1}F_{1}\left(  \frac{1}{2}\left[  1-\left\vert m+\alpha\right\vert
-\frac{k_{out}^{2}}{2M\omega}\right]  ,1-\left\vert m+\alpha\right\vert
,M\omega r^{2}\right)  \bigg\},\label{21}
\end{align}
$\ $for $r>R.$

The effect of the delta function is then taken into account by means of the
continuity relations
\begin{equation}
f_{m}(R-\varepsilon)_{in}=f_{m}(R+\varepsilon)_{out}\text{ },\label{22}%
\end{equation}%
\begin{equation}
\left[  \frac{df_{m}(r)}{dr}\right]  _{R-\varepsilon}^{R+\varepsilon}%
=\frac{\alpha s}{R}f_{m}(R-\varepsilon)_{in}\text{ }.\label{23}%
\end{equation}
In the lowest order in $R$ $\ $the confluent series and its first derivative
behave like%
\begin{equation}
\left[  _{1}F_{1}\left(  a,c,M\omega r^{2}\right)  \right]  _{r=R}%
\thickapprox1,\label{24}%
\end{equation}%
\begin{equation}
\left[  \frac{d\text{ }_{1}F_{1}\left(  a,c,M\omega r^{2}\right)  }%
{dr}\right]  _{r=R}=\frac{2a}{c}M\omega R,\label{25}%
\end{equation}
so Eq. (\ref{22}) gives
\begin{equation}
C_{m}R^{\left\vert m\text{ }\right\vert }=A_{m}R^{\left\vert m+\alpha
\right\vert }+B_{m}R^{-\left\vert m+\alpha\right\vert }.\label{26}%
\end{equation}
By applying Eq. (\ref{22}) and Eq. (\ref{23}), one obtains for the unnormalized
$f_{m}(r)$ for $r>R$ (to lowest order in $R$)
\begin{align}
f_{m}(r)  & =R^{\left\vert m\text{ }\right\vert }\bigg\{R^{-\left\vert
m+\alpha\right\vert }\text{ }\frac{\left\vert m\right\vert +\left\vert
m+\alpha\right\vert +\alpha s+R^{2}M\omega\biggl[\dfrac{(\left\vert
m\right\vert +1-\frac{k_{in}^{2}}{2M\omega})}{\left\vert m\right\vert
+1}-\dfrac{(1-\left\vert m+\alpha\right\vert -\frac{k_{out}^{2}}{2M\omega}%
)}{1-\left\vert m+\alpha\right\vert }\biggr]}{2\left\vert m+\alpha\right\vert
+R^{2}M\omega\biggl[\dfrac{-\frac{k_{out}^{2}}{2M\omega}}{\left\vert
m+\alpha\right\vert +1}-\dfrac{-\frac{k_{out}^{2}}{2M\omega}}{1-\left\vert
m+\alpha\right\vert }\biggr]}\nonumber\\
& \times r^{\left\vert m+\alpha\right\vert }\text{ }_{1}F_{1}\left(  \frac
{1}{2}\left[  \left\vert m+\alpha\right\vert +1-\frac{k_{out}^{2}}{2M\omega
}\right]  ,\left\vert m+\alpha\right\vert +1,M\omega r^{2}\right)
\nonumber\label{27}\\
& +R^{\left\vert m+\alpha\right\vert }\biggl[1-\frac{\left(  \left\vert
m\right\vert +\left\vert m+\alpha\right\vert +\alpha s\right)  +R^{2}%
M\omega\biggl[\dfrac{(\left\vert m\right\vert +1-\frac{k_{in}^{2}}{2M\omega}%
)}{\left\vert m\right\vert +1}-\dfrac{(1-\left\vert m+\alpha\right\vert
-\frac{k_{out}^{2}}{2M\omega})}{1-\left\vert m+\alpha\right\vert }%
\biggr]}{2\left\vert m+\alpha\right\vert +R^{2}M\omega\biggl[\dfrac
{-\frac{k_{out}^{2}}{2M\omega}}{\left\vert m+\alpha\right\vert +1}%
-\dfrac{-\frac{k_{out}^{2}}{2M\omega}}{1-\left\vert m+\alpha\right\vert
}\biggr]}\biggr]\nonumber\\
& \ \times r^{-\left\vert m+\alpha\right\vert }\text{ }_{1}F_{1}\left(
\frac{1}{2}\left[  1-\left\vert m+\alpha\right\vert -\frac{k_{out}^{2}%
}{2M\omega}\right]  ,1-\left\vert m+\alpha\right\vert ,M\omega r^{2}\right)
\bigg\}\exp(-\frac{M\omega}{2}r^{2}).
\end{align}
\\
By taking the limit $R\rightarrow0$, $(R{{}^2}\thickapprox0)$, one clearly obtains

\begin{align}
f_{m}(r)  & =R^{\left\vert m\text{ }\right\vert } \bigg\{R^{-\left\vert
m+\alpha\right\vert }\left(  \frac{1}{2}+\frac{\left\vert m\right\vert +\alpha
s}{2\left\vert m+\alpha\right\vert }\right)  r^{\left\vert m+\alpha\right\vert} \nonumber\\
& \times \text{ }_{1}F_{1}\left(  \frac{1}{2}\left[  \left\vert m+\alpha\right\vert
+1-\frac{k_{out}^{2}}{2M\omega}\right]  ,\left\vert m+\alpha\right\vert+1,M\omega r^{2}\right) 
+R^{\left\vert m+\alpha\right\vert }\left(  \frac{1}{2}-\frac{\left\vert
m\right\vert +\alpha s}{2\left\vert m+\alpha\right\vert }\right)
r^{-\left\vert m+\alpha\right\vert }\text{ } \nonumber\\
& \times\text{ }_{1}F_{1}\left(  \frac{1}{2}\left[  1-\left\vert
m+\alpha\right\vert -\frac{k_{out}^{2}}{2M\omega}\right]  ,1-\left\vert
m+\alpha\right\vert ,M\omega r^{2}\right)  \bigg\} \exp\left(-\frac{M\omega
}{2}r^{2}\right)  .
\end{align}

We note that the computed coefficients of the last equation are the same as in
\cite{HAGEN1}. Thus, the solution of Eq. (\ref{13}) in the limit $R\rightarrow
0$ must always be the regular one unless Eqs. (9) and (10) of \cite{HAGEN1}
are both satisfied. In this latter case only the irregular solution is allowed.

Eqs. (11.a) and (11.b) in \cite{HAGEN1} claim that the irregular solution
occurs when
\begin{align}
(a)\text{ \ }m &=-N,\text{ \ \ \ }N\geq0,\text{ \ \ \ \ \ }s=-1\text{\ \ \ or}\label{28} \\
(b)\text{ \ }m &=-N-1,\text{ \ \ }N+1\leq0,\text{ \ \ \ \ \ }s=+1,\label{29}
\end{align}
then, the wave function $\psi_{1}$ is given by
\begin{align}
\psi_{1}(r,\theta)  & =\underset{m=-\infty}{\overset{+\infty}{\sum}^{\prime}%
}a_{m}r^{\left\vert m+\alpha\right\vert }\text{ }_{1}F_{1}\left(  \frac{1}%
{2}\left[  \left\vert m+\alpha\right\vert +1-\frac{k_{out}^{2}}{2M\omega
}\right]  ,\left\vert m+\alpha\right\vert +1,M\omega r^{2}\right)
e^{-\frac{M\omega}{2}r^{2}}e^{im\theta}\nonumber\label{30}\\
& +\theta(-s)\theta(\alpha)a_{-N}e^{-iN\theta}r^{-\xi}\text{ }_{1}F_{1}\left(
\frac{1}{2}\left[  1-\xi-\frac{k_{out}^{2}}{2M\omega}\right]  ,1-\xi,M\omega
r^{2}\right)  e^{-\frac{M\omega}{2}r^{2}}\nonumber\\
& +\theta(s)\theta(-\alpha)a_{-N-1}e^{-i(N+1)\theta}r^{\xi-1}\text{ }_{1}%
F_{1}\left(  \frac{1}{2}\left[  \xi-\frac{k_{out}^{2}}{2M\omega}\right]
,\xi,M\omega r^{2}\right)  e^{-\frac{M\omega}{2}r^{2}},
\end{align}

where the prime on the summation indicates the omission of the two terms
specified by Eqs. (\ref{28}),(\ref{29}), and
\begin{equation}
\alpha=N+\xi,\text{ \ \ }0\leq\xi<1\text{ \ \ \ and }N\text{ \ is fixed
integer.}\label{31}%
\end{equation}
The confluent series behaves asymptotically at large positive values of its
argument as%
\begin{equation}
F(a,c,z)\rightarrow\frac{\Gamma(c)}{\Gamma(a)}e^{z}z^{a-c},\label{32}%
\end{equation}
so, the wave function $\psi_{1}$ behaves as%
\begin{align}
\psi_{1}  & \sim\sum_{m=-\infty}^{+\infty\prime}a_{m}r^{\left\vert
m+\alpha\right\vert }e^{\frac{M\omega}{2}r^{2}}r^{-\frac{1}{2}\left(\frac{k_{out}^{2}%
}{M\omega}+\left\vert m+\alpha\right\vert +1\right)  }\nonumber\\
& +\theta(-s)\theta(\alpha)a_{-N}e^{-iN\theta}r^{-\xi}e^{\frac{M\omega}%
{2}r^{2}}r^{-\frac{1}{2}\left(  \frac{k_{out}^{2}}{M\omega}+1-\xi\right)  } \nonumber\\
& +\theta(s)\theta(-\alpha)a_{-N-1}e^{-i(N+1)\theta}r^{\xi-1}e^{\frac{M\omega
}{2}r^{2}}r^{-\frac{1}{2}\left(  \frac{k_{out}^{2}}{M\omega}+\xi\right) },\label{33a}
\end{align}
where $\theta(x)$ is the Heaviside step function and $a_{m}$ \ is a constant.

The last expression is exponentially divergent. The divergence cannot be
avoided except by putting the parameter \cite{wang} $a=-n$, with
$n=0,1,2,....$, thus transforming the series into a polynomial of degree $n$
(Laguerre polynomials). Hence%
\begin{align}
\psi_{1}(r,\theta)  & =\underset{-\infty}{\overset{+\infty}{\sum}^{\prime}%
}\left[  a_{m,n}\text{ }r^{\left\vert m+\alpha\right\vert }L_{n}^{\left\vert
m+\alpha\right\vert }(M\omega r^{2})e^{-\frac{M\omega}{2}r^{2}}e^{im\theta
}\right] \nonumber\\ 
&+\theta(-s)\theta(\alpha)a_{-N,n\text{ }}r^{-\xi}L_{n}^{-\xi
}(M\omega r^{2})e^{-\frac{M\omega}{2}r^{2}}e^{-iN\theta}\nonumber\\
&+\theta(s)\theta(-\alpha)a_{-N-1,n\text{ }}r^{\xi-1}L_{n}^{\xi-1}(M\omega
r^{2})e^{-\frac{M\omega}{2}r^{2}}e^{-i(N+1)\theta},\label{34}%
\end{align}
where the constant $a_{m,n}=\frac{n!\Gamma(\left\vert m+\alpha\right\vert
+1)}{\Gamma(\left\vert m+\alpha\right\vert +1+n)}a_{m}$.

When the conditions (\ref{28})-(\ref{29}) are not satisfied, the energy
spectrum for the regular solution is obtained from the condition
\begin{equation}
\frac{1}{2}\left[  \left\vert m+\alpha\right\vert +1-\frac{k_{out}^{2}%
}{2M\omega}\right]  =-n,\text{ }\label{35}%
\end{equation}
thus, it follows from Eq. (\ref{35}) that%
\begin{equation}
E=\pm\sqrt{M%
{{}^2}%
+2M\omega\left[  \left\vert m+\alpha\right\vert +s\left(  m+\alpha\right)
+2n\right]  }.\label{36}%
\end{equation}
The energy spectrum when the irregular solution occurs, for the first case
($m=-N,$ \ $N\geq0,$ $s=-1)$, is deduced from
\begin{equation}
\frac{1}{2}\left[  1-\xi-\frac{k_{out}^{2}}{2M\omega}\right]  =-n\ ,\text{
\ }\label{37}%
\end{equation}
then

\begin{equation}
E=\pm\sqrt{M%
{{}^2}%
+4M\omega\left[  n-\xi\right]  }.\label{38}%
\end{equation}
For the second case ($m=-N-1,$ \ $N+1\leq0,$ $s=+1$)%
\begin{equation}
\frac{1}{2}\left[  \xi-\frac{k_{out}^{2}}{2M\omega}\right]  =-n,\text{
}\label{39}%
\end{equation}
it follows that

\begin{equation}
E=\pm\sqrt{M^{2}+4M\omega\left[  n+\xi-1\right]  }.\label{40}%
\end{equation}
We remark that the energy levels given by Eqs. (\ref{38})-(\ref{40}) depend
only on the AB parameter $\xi$ defined by Eq. (\ref{31}).

The lower component $\psi_{2}$ can be computed from Eq. (\ref{6})%
\begin{equation}
E\psi=\left[  \beta\gamma^{r}\pi_{r}+\beta\gamma^{\theta}\pi_{\theta}%
-iM\omega\beta\gamma^{r}r\beta+\beta M\right]  \psi,\label{41}%
\end{equation}
where
\begin{equation}
\gamma^{r}=\gamma^{1}\cos\theta+\gamma^{2}\sin\theta,\label{42}%
\end{equation}%
\begin{equation}
\gamma^{\theta}=-\gamma^{1}\sin\theta+\gamma^{2}\cos\theta.\label{43}%
\end{equation}
This leads to two first-order coupled differential equations
\begin{equation}
\left\{
\begin{array}
[c]{c}%
E\psi_{1}=M\psi_{1}+e^{-is\theta}\left(  \pi_{r}-is\pi_{\theta}+iM\omega
r\right)  \psi_{2}\\
E\psi_{2}=e^{is\theta}\left(  \pi_{r}+is\pi_{\theta}-iM\omega r\right)
\psi_{1}-M\psi_{2}%
\end{array}
\right. \label{44}%
\end{equation}
The second equation in the set\ of Eqs. (\ref{44}) \ leads to
\begin{equation}
\psi_{2}=\frac{-i}{(E+M)}e^{is\theta}\left(  \frac{d}{dr}-s\frac{1}{r}\frac
{1}{i}\frac{d}{d\theta}-s\frac{\alpha}{r}+M\omega r\right)  \psi
_{1}.\label{46}
\end{equation}
Using the relation
\begin{equation}
\frac{dL_{n}^{\mu}(M\omega r%
{{}^2}%
)}{dr}=-2M\omega rL_{n-1}^{\mu+1}\left(  M\omega r%
{{}^2}%
\right)  ,\text{ \ \ }n\geq1,\label{47}%
\end{equation}
and substituting Eq. (\ref{34}) into Eq. (\ref{46}) one obtains%
\begin{align}
\psi_{2}  & =\frac{-ie^{is\theta}e^{-\frac{M\omega}{2}r^{2}}}{(E+M)}%
\bigg\{\underset{m=-\infty}{\overset{+\infty}{\sum}^{\prime}}a_{m,n}%
\biggl[\left[  \mu-s\left(  \alpha+m\right)\right] r^{\mu-1} L_{n}^{\mu}(M\omega r^{2})-2M\omega r^{\mu+1}L_{n-1}^{\mu+1}(M\omega r^{2})\biggr]e^{im\theta}\nonumber\\
& +\theta(-s)\theta(\alpha)a_{-N,n}\biggl[\left[-\xi+s\left(
N-\alpha\right)\right] r^{-\xi-1}L_{n}^{-\xi}(M\omega r^{2})-2M\omega
r^{-\xi+1}L_{n-1}^{-\xi+1}(M\omega r^{2})\biggr] e^{-iN\theta}\nonumber\\
& +\theta(s)\theta(-\alpha)a_{-N-1,n}\biggl[\left[\xi-1-s\left(
N+1\right)-s\alpha\right]r^{\xi-2}L_{n}^{\xi-1}(M\omega r^{2})\nonumber\\
&-2M\omega
r^{\xi}L_{n-1}^{\xi}(M\omega r^{2})\biggr] e^{-i(N+1)\theta}\bigg\},
\end{align}

where $\mu=\left\vert m+\alpha\right\vert$ .

\section{Analysis of energy spectrum}

Let us discuss the energy spectrum corresponding to the regular solution,
i.e.
\begin{equation}
E=\pm\sqrt{M%
{{}^2}%
+2M\omega\left[  \left\vert m+\alpha\right\vert +s\left(  m+\alpha\right)
+2n\right]  }.\label{48a}%
\end{equation}
From this relation, it is clear that the energy spectrum depends on the spin
projection and the AB magnetic flux parameter. Two cases have to be distinguished:

(a) For $m+\alpha>0,$ Eq. (\ref{48a}) becomes
\begin{equation}
E=\pm\sqrt{M%
{{}^2}%
+2M\omega\left[  \left(  m+\alpha\right)  \left(  s+1\right)  +2n\right]
}.\label{49}%
\end{equation}
For $s=1,$ it is to be noted that every state with $\left(  m,n\text{
}\right)  $\ has the same energy as that of a state with $\left(  m\mp
\ell,n\pm\ell\right)  $ where $\ell$ is an integer$.$ For $s=-1$, the energy
depends only on the natural number $n$.

(b) For $m+\alpha<0,$ Eq. (\ref{48a}) becomes
\begin{equation}
E=\pm\sqrt{M%
{{}^2}%
+2M\omega\left[  \left(  m+\alpha\right)  \left(  s-1\right)  +2n\right]
}.\ \label{50}%
\end{equation}
For $s=-1,$ we observe that every state with $\left(  m,n\text{ }\right)
$\ has the same energy as that of a state with $\left(  m\pm\ell,n\pm
\ell\right)  $ $\ $\ where $\ell$ is an integer$.$ For $s=1,$ the energy
depends again only on the natural number $n$.

On the other hand, we note that the energies Eqs. (\ref{38})-(\ref{40})
corresponding to the two cases when the irregular solution occurs are not degenerated.\\

\subsection{Nonrelativistic limit}

Let us now examine the nonrelativistic limit of Eq. (11) by setting
$E$\ $=M+\epsilon$, and considering $\epsilon$ $<<M$ ,\ so that $E%
{{}^2}%
\thickapprox M$\ $^{2}+2M$ $\epsilon,$\ we obtain%
\begin{equation}
\Big\{\frac{\pi^{2}}{2M}+\frac{1}{2}M\omega%
{{}^2}%
r%
{{}^2}%
-\omega\lbrack1-s(L_{z}+\alpha)]\sigma^{3}-s\frac{e}{2M}H\sigma^{3}%
\Big\}\psi=\epsilon\psi,\label{51}%
\end{equation}
it is worth noting that Eq. (\ref{51}) corresponds to the Pauli equation of a
2D harmonic oscillator with an additional spin-orbit term. Then, it reads in
coordinates representation
\begin{equation}
\bigg\{\frac{1}{r}\frac{\partial}{\partial r}r\frac{\partial}{\partial r}%
+\frac{1}{r^{2}}(\frac{\partial}{\partial\theta}+i\alpha)^{2}-M%
{{}^2}%
\omega%
{{}^2}%
r%
{{}^2}%
+2M\epsilon+2M\omega\lbrack1-s(L_{z}+\alpha)]\sigma^{3}-\frac{\alpha s}
{R}\delta(r-R)\sigma^{3}\bigg\}\psi=0.\label{52}
\end{equation}
By using similar steps as in the relativistic case, the shifted energy
levels $\epsilon+\omega$ for the regular solution of Eq. (\ref{52}) are
\begin{equation}
\epsilon+\omega=\omega\left[  2n+1+\left\vert m+\alpha\right\vert +s\left(
m+\alpha\right)  \right]  ,\text{ \ \ with \ }n=0,1,2,\ldots\label{53}
\end{equation}

If we drop out the energy related to the spin-orbit coupling term $s\left(
m+\alpha\right)  $ from Eq. (\ref{53}), the obtained energy coincide
with the result given in the first equation of the set (54) of Ref. \cite{park}.

For spinless particles and in absence of the AB effect ($\alpha=0$), we find
the energy levels of the circular oscillator \cite{Flugge} shifted by constant
value $\omega.$

For the two cases when the irregular solution occurs, we find
\begin{equation}
\epsilon+\omega=\omega\left[  2n+1-\left\vert m+\alpha\right\vert +s\left(
m+\alpha\right)  \right]  ,\text{ }\label{53a}%
\end{equation}
this result can be written explicitly in term of the parameter $\xi$ as%
\begin{equation}
\epsilon+\omega\ =\omega\left[  2n+1-2\xi\right]  ,\text{ \ \ \ for
\ }m=-N\ \ ,\text{ \ }N\geq0\text{ \ and \ }s=-1,\text{ }\label{54}%
\end{equation}
\begin{equation}
\epsilon+\omega=\omega\left[  2n-1+2\xi\right]  ,\text{ \ for \ }m=-N-1,\text{
\ }N+1\leq0\ \text{and }s=+1,\label{55}%
\end{equation}
where $0\leq\xi<1.$

It is interesting to note that if we drop out the energy related to the
spin-orbit coupling term $s\left(  m+\alpha\right)  $ from Eq. (\ref{53a}), 
the obtained energy coincide with the result given in the second
equation of the set (54) of Ref. \cite{park}, i.e.%
\begin{equation}
\epsilon+\omega\ =\omega\left[  2n+1-\left\vert m+\alpha\right\vert \right]
,\text{ \ \ \ }\label{56}%
\end{equation}
which is valid only for \ ($m=-N\ \ ,$ \ $N\geq0$ \ and \ $s=-1$)\ or for
\ ($m=-N-1,$ \ $N+1\leq0\ $and $s=+1$). Eq. (\ref{56})\ can be also
written explicitly in term of the parameter $\xi$ as%
\begin{equation}
\epsilon+\omega\ =\omega\left[  2n+1-\xi\right]  ,\text{ \ \ \ for
\ }m=-N\ \ ,\text{ \ }N\geq0\text{ \ and \ }s=-1,\text{ }%
\end{equation}
\begin{equation}
\epsilon+\omega=\omega\left[  2n+\xi\right]  ,\text{ \ for \ }m=-N-1,\text{
\ }N+1\leq0\ \text{and }s=+1.
\end{equation}

\section{Conclusion}
In this paper, we have found the bound states and corresponding eigenfunctions
of a 2D Dirac oscillator in the presence of the Aharonov-Bohm effect . The
energy levels depends on the spin of particle and the AB magnetic flux
parameter. It is shown that the energy spectrum can takes particular values
due to the singular solution. These values (for fixed spin projection)
depend only on the AB magnetic parameter $\xi$. Both energy levels of regular
and irregular solutions may indicate more information on the role the magnetic
vector potential. The nonrelativistic limit is also deduced. \ \ 

"The calculations presented here is another detection of an important role for
the solutions of the wave equation which are singular at the origin. It is in
fact very tempting to make the general observation that the rejection of
singular wave functions basing on their singularity should be viewed with some
suspicion" \cite{HAGEN3}.\\

\textbf{{Acknowledgments}}\newline
The authors thank Professor C. R. Hagen (University of Rochester, NY) for his
remarks and comments.


\newpage

\begin{thebibliography}{99}                                                                                             
\bibitem {Moushinsky}M. Moshinsky and A.Szczepaniak, J. Phys. \textbf{A 22},
(1989) L817.
\bibitem {MoushinskyB}M. Moshinsky and Y. F. Smirnov, \textit{The Hamonic
oscillator in Modern Physics}, Harwood Academic Publishers, Amsterdam, 1996.
\bibitem {rozmej}P. Rozmej and R. Arvieu, J. Phys. \textbf{A 32}, (1999) 5367, arXiv: quant-ph/9903073.
\bibitem {pacheco}M. H. Pacheco, R. R. Landim and C. A. S. Almeida, Phys.
Lett. \textbf{A 311}\,(2003) 93, arXiv: hep-th/0303142.
\bibitem {Vil2}V. M. Villalba and A. R. Maggiolo , Eur. Phys. J. \textbf{B 22}
(2001) 31, arXiv: cond-mat/0107529.
\bibitem {Vil3}V. M. Villalba and R. Pino, Mod. Phys. Lett. \textbf{B 17}
(2003) 1331 , arXiv: cond-mat/0402348, and references therein.
\bibitem {aharo}Y. Aharonov and D. Bohm, Phys. Rev. \textbf{115 }(1959) 485.
\bibitem {yasha}Ya-Sha Yi, Tie-Zheng Qian and Zhao-Bin Su, Phys. Rev. \textbf{B 55} (1997) 10632.
\bibitem {HAGEN1}C. R. Hagen, Phys. Rev. Lett. \textbf{64} (1990) 503.
\bibitem {Vil1}V. M. Villalba, Phys. Rev. \textbf{A 49} (1994) 586, arXiv: hep-th/9310010.
\bibitem {HAGEN2}C. R. Hagen, Int. J. Mod. Phys. \textbf{A 6} (1991) 3119.
\bibitem {wang}Z. X. Wang and D. R. Duo, \textit{Special Functions}, World
Scientific Pub., Singapore, 1989, p.327.
\bibitem {park}D. K. Park, J. Math. Phys. \textbf{36} (1995) 5453.
\bibitem {Flugge}S. FL\"{u}gge, \textit{Practical Quantum Mechanics}, 2nd ed.,
Springer, Berlin,1994, pp.107-110 (Vol. I).
\bibitem {HAGEN3}C. R. Hagen, \textit{Spin and the Aharonov-Bohm Effect},
Preprint UR-1145 (1990).
\end{thebibliography}
\end{document}